\documentclass[pre,aps,showkeys,showpacs,superscriptaddress,twocolumn,%
amsmath,amssymb]{revtex4} 

\usepackage{graphics}
\usepackage{epsfig}
\usepackage{amsmath}
\usepackage{setspace}
 
\begin{document}

\title{Classical Heisenberg spins on a hexagonal lattice with Kitaev couplings}

\author {Samarth Chandra}
\email{samarth@rri.res.in}
\altaffiliation{Present address : Raman Research Institute, C. V. Raman Avenue, Sadashivanagar, Bangalore-560080, India} 
\affiliation{Department of Theoretical Physics, Tata Institute of Fundamental Research, Homi Bhabha Road, Colaba, Mumbai-400005, INDIA}

\author {Kabir Ramola} 
\email{kabir@tifr.res.in}
\affiliation{Department of Theoretical Physics, Tata Institute of Fundamental Research, Homi Bhabha Road, Colaba, Mumbai-400005, INDIA}

\author {Deepak Dhar} 
\email{ddhar@theory.tifr.res.in}
\affiliation{Department of Theoretical Physics, Tata Institute of Fundamental Research, Homi Bhabha Road, Colaba, Mumbai-400005, INDIA}

\date{\today}

\begin{abstract}
We analyse the low temperature properties of a system of classical Heisenberg spins on a hexagonal lattice with Kitaev couplings. For a lattice of $2N$ sites with periodic boundary conditions, the ground states form an $(N+1)$ dimensional manifold. We show that the ensemble of ground states is equivalent to that of a solid-on-solid model with continuously variable heights and nearest neighbour interactions, {\it at a finite temperature}. For temperature $T$ tending to zero, 
all ground states have equal weight, and there is no order-by-disorder in this model. We argue that the bond-energy bond-energy correlations at distance $R$ decay as $\frac{1}{R^2}$ at zero temperature. This is verified by Monte Carlo simulations. We also discuss the relation to the quantum spin-S Kitaev model for large $S$, and obtain lower and upper bounds on the ground state energy of the quantum model.
\end{abstract}


\maketitle

\section{Introduction}


There has been a lot of interest in the Kitaev model in recent years. The quantum mechanical spin-1/2 Kitaev model is exactly solved in 2D. The Hamiltonian can be diagonalized exactly in terms of Majorana fermions \cite{kitaev}. The model exhibits a phase transition from a phase with finite correlation length to one with long-range correlations as the ratios of coupling constants in different directions are varied \cite{feng,lahtinen}. It has topological excitations, and their robustness with respect to noise makes it an interesting candidate for quantum computing \cite{nayak}. 

In a recent very interesting paper, Baskaran et al. studied a generalization of this model with spin-$S$ at each site and identified mutually commuting $\mathbb{Z}_2$ variables that are constants of motion for arbitrary $S$ \cite{baskaran}. For large $S$, the spins can be approximated as classical $O(3)$ vector spins. Baskaran et al. showed that the classical ground state of the model has a large degeneracy. They argued that though a naive averaging over these ground state configurations would suggest that the system is disordered at zero temperature, for large $S$, the quantum fluctuations of spins have lower energy for a subset of the classical ground states. These states get more weight in the quantum mechanical ground state, and the quantum model shows long-range order in the ground state, an example of quantum order-by-disorder.

The finite-temperature fluctuations in the classical model behave qualitatively like the zero-point fluctuations in the quantum model, and it is interesting to ask if temperature fluctuations can induce order-by-disorder in the classical Heisenberg spins with Kitaev couplings, just as the quantum fluctuations are expected to, in the large-$S$ quantum model studied by Baskaran et al.. In this paper we point out that there is a qualitative difference between the classical and quantum mechanisms of order-by-disorder.
For the classical Kitaev model, the contribution of nearby states to the restricted partition function in the limit of very small temperature, with states summed only over the neighbourhood of a given classical ground state, is exactly the same for almost all ground states. For the ensemble of ground states, we establish an exact equivalence to a solid-on-solid model with nearest neighbour coupling at a finite temperature. We argue that bond-energy bond-energy correlations decay as $R^{-2}$ with distance $R$ at zero temperature. Our Monte Carlo simulations also support this conclusion. We derive upper and lower bounds on the ground state energy of the quantum spin-$S$ model. We also study the ground state energy of the quantum system using variational wavefunctions, avoiding the divergences present in the spin-wave expansion. 

Kitaev's pioneering work has led to  a large amount of further research. 
The spectrum of the different phases of this model have been extensively studied \cite{lahtinen2}. Proposals for experimentally realising this model using polar molecules and ultracold atoms trapped in optical lattices have recently been made \cite{micheli,duan}. Alternate methods of solving this model using Jordan-Wigner transformations have been proposed \cite{feng, Kells3}, while perturbative studies have also proved fruitful \cite{vidal, dusuel}. The Kitaev model has also been studied on other two dimensional lattices \cite{karimipour, kells, sarma, Kells2, yao,yang2}. Exact solutions have been obtained for the Kitaev model on certain three dimensional lattices \cite{si, mandal}.
There is also a fair amount of earlier work on order-by-disorder in classical systems \cite{villain, moessnerchalker}. It has been studied a lot in the context of magnetic systems with frustration, such as spin systems with nearest neighbour antiferromagnetic interactions on different lattices\cite{henley1,gvozdikova}.


\section{Definition of the Model}

\begin{figure}
\centering
\includegraphics[width=1.0\columnwidth,angle=0]{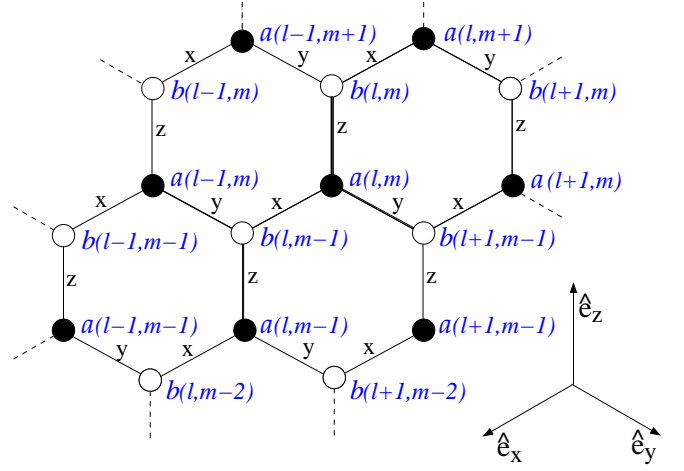}
\caption{(Color online) A hexagonal lattice depicting the labelling scheme for sites, and the x, y and z bond classes. Sites in the A- and B- sublattices are denoted by filled and open circles respectively.}
\label{hex_lat}
\end{figure}

We consider classical Heisenberg spins on a hexagonal lattice. We consider a finite lattice, with periodic boundary conditions. There are $L$ hexagons in each row, and $M$ rows of hexagons ($L$ and $M$ both assumed even). The total number of hexagons is $N=LM$, and the number of sites is $2N$. The bonds of the lattice are divided into three classes, $X,Y$ and $Z$, according to their orientation (Fig. \ref{hex_lat}). The hexagonal lattice consists of two sublattices denoted by A and B. We label sites in the A sublattice by $a(l,m)$ and the corresponding B sublattice site connected to it via a $Z$ bond by $b(l,m)$. We define three bond vectors $e_x$, $e_y$, $e_z$ as the vectors from any A site to its three neighbours via the $X,Y$ and $Z$ bonds respectively (Fig. \ref{hex_lat}). Thus we have $a(l,m)+e_x = b(l,m-1)$, $a(l,m)+e_y = b(l+1,m-1)$ and $a(l,m)+e_z = b(l,m)$. We define the hexagonal plaquette $(l,m)$ to be the hexagon whose topmost point is $a(l,m)$. A bond will be specified by the $(l,m)$ coordinate of its A-lattice end point and its class $X$, $Y$ or $Z$. For instance, $(a(l,m);x) \equiv (l,m;x)$ is an X-bond with $a(l,m)$ at one of its ends. The periodic boundary conditions are implemented by making $a(l,m) = a(l+L,m) = a(l-\frac{M}{2}, m+M)$.

At each lattice site $i$ there is a three dimensional vector spin $\vec S_{i} = ({S_i}^x,{S_i}^y,{S_i}^z)$ of unit magnitude. Thus ${{S_i}^x}^2+{{S_i}^y}^2+{{S_i}^z}^2 = 1$ at every site. The Hamiltonian of the system is given by
\begin{equation}
H = - J \sum_{a \in A} [ S_{a}^{x} S_{a +e_x}^{x} + S_{a}^{y} S_{a +e_y}^{y} + S_{a}^{z} S_{a +e_z}^{z} ] 
\label{def_Hamiltonian}
\end{equation}
As the hexagonal lattice is bipartite, for classical spins, without loss of generality we take $J > 0$. The Hamiltonian does not have rotational symmetry in the spin space, but it has a local symmetry : for any bond $(l,m;\alpha)$, the Hamiltonian is invariant under the transformation $S_{a(l,m)}^{\alpha} \rightarrow -S_{a(l,m)}^{\alpha}, S_{a(l,m)+e_{\alpha}}^{\alpha} \rightarrow -S_{a(l,m)+e_{\alpha}}^{\alpha}$.

\section{Finite temperature Partition Function} 

The partition function of the system at finite temperature $Z[\beta]$ is given by 

\begin{equation} 
Z[\beta] = \int \prod_{s} (\frac{d\overrightarrow{S_s}}{4 \pi}) 
~\textmd{exp}[-\beta H] 
\end{equation}
where $\beta^{-1} = T$. The index $s$ runs over all sites of the lattice. The integral over each B-site is of the form $W_{l,m}=\int d\overrightarrow{S}_{b(l,m)} \exp [-\beta 
\overrightarrow{S}_{b(l,m)}.\overrightarrow{F}]$ where $\overrightarrow{F} = 
S_{a(l,m+1)}\hat{i} + S_{a(l-1,m+1)}\hat{j} + S_{a(l,m)}\hat{k} $. This can in turn be evaluated as 
\begin{align}
\nonumber
& W_{l,m} = \frac{1}{2}\int_{-1}^{1} d(\cos~\theta) \\
& \exp[-\beta \cos \theta
\sqrt{{S_{a(l,m+1)}^{x}}^{2}+{S_{a(l-1,m+1)}^{y}}^{2}+{S_{a(l,m)}^{z}}^{2}}],
\end{align}
where $\theta$ is the angle between the vector $\overrightarrow{S}_{b(l,m)}$ and $\overrightarrow{F}$. This immediately yields 
\begin{equation}
 W_{l,m} = \frac{\sinh [\beta 
\sqrt{{S_{a(l,m+1)}^{x}}^{2}+{S_{a(l-1,m+1)}^{y}}^{2}+{S_{a(l,m)}^{z}}^{2}}]}{\beta 
\sqrt{{S_{a(l,m+1)}^{x}}^{2}+{S_{a(l-1,m+1)}^{y}}^{2}+{S_{a(l,m)}^{z}}^{2}}}.
\end{equation}
Now
\begin{equation} 
{Z[\beta]} =\int \prod_{(l,m)} \frac{d \overrightarrow{S}_{a(l,m)}}{4 \pi} \prod_{(l,m)} 
W_{l,m}.
\end{equation}

We thus obtain the effective Hamiltonian for the spins on the $A$-sublattice alone as
 
\begin{align}
\nonumber
& H_{eff}(\{{\vec S}_a\},\beta) = \\
&-\frac{1}{\beta}\sum_{(l,m)} 
F(\beta(\sqrt{{S_{a(l,m+1)}^{x}}^{2}+{S_{a(l-1,m+1)}^{y}}^{2}+{S_{a(l,m)}^{z}}^{2}})) ,
\label{eq:heff}
\end{align}
where 
\begin{equation}
 F(x) = \log[\frac{\sinh (x)}{x}] 
\label{def_Fx}.
\end{equation}

For a given configuration of spins $\{S^{\alpha}\}$, to each bond $(l,m;\alpha)$ of the lattice, we assign a vector $\epsilon(l,m;\alpha) {\vec{e}}_{\alpha}$, with $\epsilon(l,m; \alpha)$ given by 

\begin{equation}
\epsilon(l,m;\alpha) = (S^{\alpha}_{a(l,m)})^2 - \frac{1}{3}.
\label{GS_condition}
\end{equation}

We define the discrete divergence of the $\epsilon$-field on the A, and B sublattices as
\begin{align}
&\nabla. \epsilon \equiv \sum_{\alpha}\epsilon(a(l,m) ;\alpha) {\rm ~at ~site} ~a(l,m),\\
& \nabla. \epsilon \equiv \sum_{\alpha}\epsilon(b(l,m) - e_{\alpha};\alpha) {\rm ~at ~site} ~b(l,m).
\end{align}

Clearly, the divergence of the field $\epsilon$ at any site on the A-sublattice is $0$. In addition, {\it for ground state configurations} ( the proof of this assertion is given the next section), even for bonds meeting at any site, $b \in B$, we have 

\begin{equation}
\sum_{\alpha}\epsilon(b(l,m)- e_{\alpha};\alpha) = 0, {\rm ~for ~all~sites~} b \in B . 
\label{gs_constraints}
\end{equation}

For non-ground state configurations, the sum of $\epsilon$ variables at each $B$-site is no longer unity. We parametrise this deviation using the new variable $Q$ defined by
\begin{equation}
 Q_{b(l,m)} = - \sum_{\alpha} \epsilon(b(l,m) - e_{\alpha}; \alpha). 
\end{equation}

We can think of these variables $Q_{b(l,m)}$ as charges which are placed on the sites of the $B$-sublattice; there are no charges on the $A$-sites. We define $Q_{a(l,m)} =0$, for all sites $a(l,m)$.
Given the values of these charges, we can construct the corresponding electrostatic potential field $\phi$ defined at all sites $s$ of the lattice such that 
\begin{equation}
\nabla^2 \phi(s) = -Q_{s}; {\rm ~ for ~ all ~ sites ~} s.
\end{equation}

\begin{figure}[h]
\centering
\includegraphics[width=.7\columnwidth,angle=0]{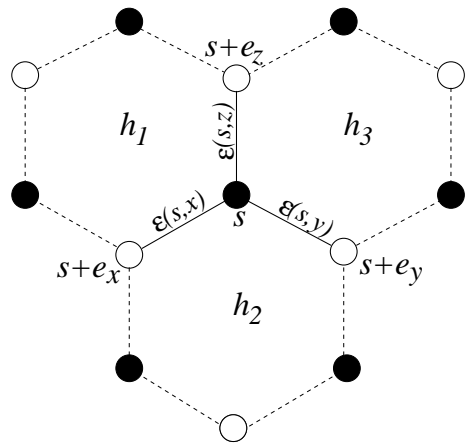}
\caption{Figure depicting the definition of the $\epsilon$ and $h$ variables on the bonds and plaquettes around an $A$-site $s$.}
\label{fig_epsilon_def}
\end{figure}

Here $\nabla^2$ is the discrete laplacian on the lattice. These equations can be solved explicitly, 
and determine the potential $\phi(s)$ completely, up to an overall additive constant, so long as the total charge in the system is zero. Explicitly, we have
\begin{equation}
\phi(s) = \sum_{s'} G(s,s') Q_{s'},
\label{eqphi}
\end{equation}
where $G(s,s')$ is the lattice Green's function. Then, as $\nabla^2 \phi (s) = - Q_s = - \nabla. \epsilon (s)$, we see that $ \epsilon + \nabla \phi $ has no divergence and can be expressed in terms of the curl of a new field $f$. We define the scalar field $f(l,m) \equiv f_{l,m}$ attached to the hexagons of the lattice (as shown in Fig. \ref{fig_epsilon_def}) such that the difference in the $f$ field between two neighbouring plaquettes is equal to the value of $ \epsilon + \nabla \phi $ along the shared bond. This satisfies the divergence-free condition for the field $ \epsilon + \nabla \phi $. Let $s$ be any site on the $A$-sublattice. Its neighbours are sites $s + e_x, s+ e_y, s+e_z$. Let the three hexagons to which $s$ belongs be $h_1, h_2$ and $h_3$ (Fig. \ref{fig_epsilon_def}).
If the site $s \equiv a(l,m)$, then $h_1$ will have coordinates $(l-1,m+1)$, and similarly for other hexagons. Then, for all sites $s$, the $f$- field is defined by
\begin{eqnarray}
\nonumber
\epsilon(s,x) = \phi(s) - \phi(s + e_x) + f(h_1)- f(h_2) \\
\nonumber
\epsilon(s,y) = \phi(s) - \phi(s + e_y) + f(h_2)- f(h_3) \\
\epsilon(s,z) = \phi(s) - \phi(s +e_z) + f(h_3)- f(h_1) 
\label{eq:def-epsilon}
\end{eqnarray}

Given the fields $\epsilon(s,\alpha)$ and $ \phi(s)$, we assign any fixed value to $f(l,m)$ at one particular hexagon, then the value of the $f$-field at neighbouring hexagons is completely determined. Thus for a given configuration of $\{S^{\alpha}\}$, we can determine the $f$-fields at all hexagons up to an overall additive constant.

We will use the values of $\{Q_s\}$ and $\{f(l,m)\}$, instead of $\{{S^{\alpha}}^2\}$ to specify the spin-configurations. The number of variables $ {S^{\alpha}_s}^2$ are $3 N$ in number, with $N$ constraints between them, thus there are $2 N$ independent real variables. 
As the variables $Q_s$ satisfy the constraint $\sum_s Q_s = 0$, there are $N-1$ independent parameters $Q_s$. Also, there are only $(N-1)$ independent parameters $f(l,m)$, as these are defined only up to an overall additive constant. We need two additional linearly independent variables to complete our new set of coordinates. We choose these to be $R_1= \sum_{l} \epsilon(l,m;z)+ \phi(b(l,m))-\phi(a(l,m))$ and $R_2= \sum_{m} \epsilon(l,m;y) + \phi(b(l+1,m-1))-\phi(a(l,m))$, which lead to
 
\begin{eqnarray}
\nonumber
f(l +L,m) = f(l,m) + R_1\\
f(l-\frac{M}{2},m+M) = f(l,m) + R_2
\label{boundarycondition}
\end{eqnarray}

Assuming that the $\phi$- field, obtained in Eq.(\ref{eqphi}) is periodic on the torus,
with $\phi(a(l,m)) = \phi(a(l+L,m)) = \phi(a(l-M/2,m+M))$, $R_1$ is independent of $m$ (and $R_2$ of $l$), and these correspond to fixing the boundary conditions for the $f(l,m)$.

The condition $2/3 \geq \epsilon(s, \alpha) \geq -1/3$ at each bond implies constraints on the allowed range of $f(l,m)$ and $Q(l,m)$. Now, given the values of the $Q$, $f$ and $\phi$ fields, one can systematically reconstruct the $\epsilon$ field (and thus the spin configuration). The value of $\epsilon$ at any bond in the bulk can be evaluated from the $f$'s at the neighbouring plaquettes and the $\phi$ at the end of the bond as shown in Eq. (\ref{eq:def-epsilon}). The $\epsilon$'s at the edges of the lattice are determined by the values of $f$ at the plaquettes next to the edge, which can be obtained from $R_{1}$ (or $R_{2}$) using Eq. (\ref{boundarycondition}). Thus, in the allowed range, the transformation from $\{Q,f,R \}$ to $ \{ {S^{\alpha}}^2 \}$ is invertible. We can now express the partition function in terms of these new variables.

Firstly we analyse the phase space factors in the partition function as we change variables from \{$\vec{S}$\} to \{$f,Q$\}. The phase space integral for each $A$-site spin is
\begin{equation}
\int \vec{dS} = \int d{S^{x}} d{S^{y}} d{S^{z}} \delta ({S^{x}}^2 + {S^{y}}^2 + {S^{z}}^2 -1).
\end{equation}

We change our integration variables from $S_s^x$ to $(S_s^x)^2$. We have $dS_s^x = d(S_s^x)^2/ (2 \sqrt{ (S_s^x)^2})$ and similarly for the y and z components. The $f$'s and $Q$'s are linear functions of the $(S^{\alpha}_s)^2$'s ($\alpha = x, y, z$), hence the Jacobian matrix of transformation for this change of variables is a $2N \times 2N$ constant matrix. Also, the determinant is non-zero as the transformation is invertible. The partition function at finite temperature, up to an unimportant constant, is thus given by 
{\small
\begin{align}
\nonumber
& Z[\beta]= (\textmd{Const.})\left[ \int dR_1 dR_2 \right] \left[\prod_{l,m} \int df_{l,m} \int {dQ_{b(l,m)}}\right] \\ 
& \left[ \prod_{\textmd{bonds}} \left( \frac{1}{3} + \epsilon(\textmd{bond}) \right)^{-1/2} \right] \textmd{exp}\left[{\sum_{l,m} F(\sqrt{1 + Q_{b(l,m)}})}\right],
\end{align}
}

where, $\Pi_{bonds}$ denotes the product over all bonds $(l,m,\alpha)$ of the lattice, with 
the $\epsilon$ variables defined in Eq. (\ref{GS_condition}), and $F(x)$ is defined by Eq. (\ref{def_Fx}).

\section{Characterisation of the ground state manifold}

Baskaran et al. defined Cartesian states as states where each spin is aligned in the direction along a Cartesian axis (x-, or y- or z-) \cite{baskaran}. We can construct a Cartesian ground state of $H$, by constructing a dimer covering of the hexagonal lattice. Spins at the ends of a dimer of type $\alpha$ ($\alpha = X, Y $ or $Z$), are aligned parallel to each other in the direction $\alpha$ (either both having $S^{\alpha} = +1$, both having $S^{\alpha} = -1$). This state has an energy $ -NJ$. Then corresponding to a dimer covering, there are $2^N$ Cartesian ground states. The number of dimer coverings of the hexagonal lattice increases as 
$1.38^N$ \cite{kasteleyn}, hence the number of Cartesian ground states increases as $ 2.76^N$.
Baskaran et al. also showed that for any two Cartesian states, there is a one-parameter family of ground states that connects them, thus forming a network of ground states.

However, they did not provide a proof that no states of lower energy can be formed, or study other possible ground states. In this section we characterise the entire set of ground states of this model.

In the large $\beta$ limit, F(x) can be replaced by $x$. Therefore the ground state energy $E_0$ of the system is given by 

\begin{equation}
E_0 = -J ~ \textmd{Max} \left[ \sum_{(l,m)} \sqrt{ 1 + Q_{(l,m)}} \right].
\label{ineq}
\end{equation}

As $\sqrt{x}$ is a convex function of $x$, for all real positive $x_i$, $ i = 1$ to $N$, we have

\begin{equation}
- \sum_i \sqrt{x_i} \geq N \left( - \sqrt{\frac{\sum_i x_i}{N}} \right)
\label{convex_ineq}
\end{equation}
With equality holding only when all the $x_{i}$ are equal. Using this inequality in (\ref{ineq}), we get

\begin{equation}
E_{min}(\{\vec S_{a}\}) \geq - JN. 
\end{equation}

This result can also be arrived at by noting that $F[\beta \sqrt{x}]$ is a convex function for all $\beta , x > 0$. As the equality sign in Eq. (\ref{ineq}) holds only when all the terms within the square root are equal, the necessary and sufficient condition for the ground state configuration is 
\begin{equation}
Q_{b} =0, {\rm ~for ~all~ sites~} b \in B.
\end{equation}

Since the $Q$ -field, and hence also the $\phi$-field are exactly zero everywhere in the ground states, the manifold is described only by the $f$-field. Correspondingly, the equations (\ref{eq:def-epsilon}) simplify to
\begin{eqnarray}
\nonumber
\epsilon(s,x) = f(h_1)- f(h_2) \\
\nonumber
\epsilon(s,y) = f(h_2)- f(h_3) \\
\epsilon(s,z) = f(h_3)- f(h_1) 
\label{eq:def-epsilon0}
\end{eqnarray}

The set of states forms an $N+1$ dimensional manifold, parametrized by the variables $\{f\}$, with the boundary conditions on these given by $R_1$ and $R_2$. It is a convex set whose extremal points correspond to the Cartesian states studied by Baskaran et al.. 

We define the restricted partition functions for a fixed $\{f(l,m)\}$ by integrating over the $Q$'s
{\small
\begin{equation}
Z\left[\{f_{(l,m)}\},\beta\right] = \left[ \prod_{(l,m)} \int dQ_{b(l,m)} \right] \textmd{exp}[ -\beta H_{eff}[ \{f,Q\}]. 
\label{restricted_part_fn}
\end{equation}
}

For large $\beta$, the integrand in Eq.(\ref{restricted_part_fn}) is sharply peaked at $Q_s=0$. We can use the method of steepest descent to find the value of this integral (integrating over the $N-1$ $Q$ variables).
We expand $ H_{eff}$ in a power series in $Q$'s
\begin{align}
H_{eff} = E_{0} + \sum_{s} {Q_s}^2 + \ldots
\end{align}
The linear term in $Q$ in the function $H_{eff}$ vanishes since the $\sum_s Q_s =0$. While the range of the $Q_s$ integrals depend on \{$f_{l,m}$\}, for large $\beta$, when the width of the peak is much smaller than the range of integration, and the peak is away from the end points of the range, each integration to leading order is independent of \{$f_{l,m}$\} and gives a factor $C \beta^{-1/2}$ where $C$ is a constant. The restricted partition function $Z[\{f\},\beta]$ in the limit of very small temperature to leading order in $\beta$, is $\beta^{-(N-1)/2} Z_{0}[\{f\}]$. Where $Z_{0}[\{f\}]$ is given by

\begin{align}
\nonumber
& Z_{0}[\{f\}] = \lim_{ \beta \rightarrow \infty} \beta^{(N-1)/2} Z[\{f\},\beta] \\ 
& = \textmd{Const.} \left[ \prod_{\textmd{bonds}} [ \frac{1}{3} + \epsilon(\textmd{bond})]^{-1/2} \right].
\label{zero_temp_part_fn}
\end{align}
Thus for all fixed $\{f(l,m)\}$, the integration over fluctuations in $\{Q\}$ produces the same temperature dependent weight factor, in the limit of large $\beta$. For evaluating averages in the limit of low temperatures, we can ignore the $Q$-degrees of freedom, and set them equal to zero. Now, the zero-temperature partition function, i.e.- the partition function in the limit $\beta \rightarrow \infty$ defined as 
\begin{align}
\nonumber
& Z_0 = \lim_{ \beta \rightarrow \infty} \beta^{(N-1)/2} Z[\beta] \\
& = \left[\int dR_1 dR_2\right]\left[ \prod_{(l,m)} \int df(l,m) \right] Z_{0}[\{f\}]
\end{align}
can be expressed as
{\small
\begin{equation}
Z_0 = \left[\int dR_1 dR_2\right] \left[ \prod_{(l,m)} \int df(l,m) \right] \textmd{exp} \left[ -\sum_{\langle i,j \rangle} V(f_i -f_j) \right],
\end{equation}
}
where 
\begin{eqnarray}
\nonumber
V(x) &=& \frac{1}{2} \log ( \frac{1}{3} + x), {\rm ~for ~} -1/3 \leq x \leq +2/3;\\
&=& +\infty, {\rm otherwise}.
\end{eqnarray}
and the sum over $\langle i,j \rangle$ denotes the summation over all nearest neighbour hexagons $i$ and $j$.


\section{ Equivalence to the solid-on-solid model}

We note that $Z_{0}$ may be interpreted as the partition function of a solid-on-solid (SOS) model, with a real height variable $f_{l,m}$ located at sites $(l,m)$ of a triangular lattice and interacting via an effective Hamiltonian $H_{SOS}$. This Hamiltonian depends on the temperature $T$ of the spin model. At a finite temperature $H_{SOS}$ is determined by integrating over the $Q$ variables in the restricted partition in Eq. (\ref{restricted_part_fn}). $H_{SOS}(T>0)$ has some weak long range couplings. However at $T=0$ the ${Q}$ field is identically zero, which leads to a purely nearest-neighbour, but non-quadratic coupling between the height variables given by 
\begin{align}
\nonumber
& H_{SOS}(T=0) = - \sum_{(l,m)} [ V(f_{l,m-1} - f_{l-1,m})\\& + V(f_{l,m} -f_{l,m-1})
+ V(f_{l-1,m} -f_{l,m})].
\end{align}

We note that the Hamiltonian has a term $\log(\frac{1}{3} + \epsilon(\textmd{bond}))$, which diverges when $\epsilon(\textmd{bond})$ tends to $-1/3$. Thus the Cartesian states of Baskaran et al. have a large relative weight, which has a divergent density. However, this divergence is an integrable divergence, and the actual measure of the Cartesian states in the ensemble of states at zero temperature is zero.

The gauge symmetry of the model has a consequence that all correlation functions of the type $\langle S_{s1}^{\alpha} S_{s2}^{\beta} \rangle$ with sites $s_1$ and $s_2$ not nearest neighbours are zero \cite{baskaran2}. The simplest nontrivial correlation functions, for non-neighbour $s_1$ and $s_2$ are of the type $\langle (S_{s1}^{\alpha})^2 (S_{s2}^{\beta})^2 \rangle$. The convergence of the high temperature expansion of the partition function implies that these correlations fall exponentially with distance, at small $\beta$. As there is no phase transition as $\beta \rightarrow \infty$, we expect this behaviour for all $0 \leq \beta < \infty$ as well, with the correlation length increasing as a function of $\beta$.

We now argue that, at zero temperature, this correlation function decays as $R^{-2}$ for large separations $R$.

The SOS model has the symmetry that changing all heights by the same constant leaves the Hamiltonian unchanged. Though the interaction is a strongly non-linear function of $f_{b(l,m)+e_{\alpha}} -f_{b(l,m)+e_{\alpha'}}$, one expects that in the high-temperature phase of the SOS model, the long-wavelength hydrodynamical modes in the system will still be sound-like, with effective Hamiltonian $|{\nabla f}|^{2}$, which gives rise to the spectrum given by $\omega^2 \propto k^2$. For two sites $s_1$ and $s_2$ separated by a large distance $R$

\begin{equation}
\langle(f_{s_{1}} - f_{s_{2}})^2\rangle \sim \log R.
\end{equation}

This implies that 
\begin{equation}
\langle\nabla f_{s_{1}} . \nabla f_{s_{2}}\rangle \sim \frac{1}{R^2}.
\end{equation}

Since the energy density variables $(S_s^{\alpha})^2$ are proportional to $\nabla f$ (at zero temperature), we conclude that the connected part of the bond-energy bond-energy correlation function 

\begin{equation}
\langle(S^{\alpha}_{s_1})^2 (S^{\beta}_{s_2})^2\rangle_c~~ \sim ~~ \frac{1}{R^2}.
\label{powerlaw}
\end{equation}

At infinite temperature, the spins at different sites are completely uncorrelated. This is not true for the $f$ variables, which have non-trivial correlations even for $\beta = 0$. In the Appendix we calculate the leading behaviour of $\langle(f_{R}-f_{0})^{2}\rangle$ at large $R$ for $\beta = 0$. We have, at infinite temperature
\begin{equation}
\langle(f_{R}-f_{0})^{2}\rangle_{\beta=0} = \frac{2 \sqrt{3}}{45 \pi} \textmd{log}[R] + {\cal O}(1) ~~~~\textmd{for large R}.
\end{equation}


\section{ Monte-Carlo simulations}

In this section we present results from Monte Carlo studies of this model for the zero temperature as well as for non-zero temperatures. 

We simulated the effective Hamiltonian $H_{eff}$ (Eq. \ref{eq:heff}), obtained by integrating out spins on the $B$-sublattice. For the finite temperature simulations, two kinds of moves were employed---single spin moves and cluster moves.

We discuss single spin moves first. In any given state, we choose a site, $a(l,m)$. We generate a gaussian random vector $\vec{r}=(r_x, r_y, r_z)$, whose variance is proportional to the temperature $T$. The proposed single spin move is then to change the spin at site $a(l,m)$ from ${\vec S_{a(l,m)}}$ to $\vec{S}_{a(l,m)}'$, given by 

\begin{equation}
\vec{S}_{a(l,m)}' = \frac{\vec{S}_{a(l,m)} + \vec{r}_s}{|\vec{S}_{a(l,m)} + \vec{r}_s|}.
\end{equation}

If the change in the effective Hamiltonian by the move is $ \Delta H$, the move is accepted with probability Min$( 1, e^{ -\beta \Delta H})$. Clearly, this satisfies the detailed balance condition.

\begin{figure}
\centering
\includegraphics[width=1.0\columnwidth,angle=0]{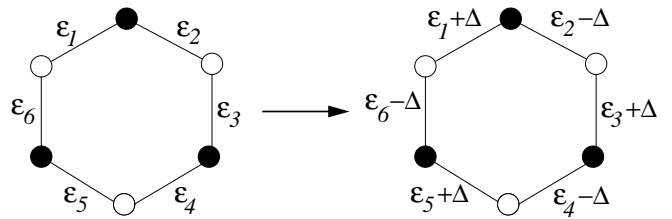}
\caption{The hexagon update move in the Monte Carlo simulations. The $\epsilon$'s depict the bond-energy variables associated with each A-site (depicted by filled circles). A random number uniformly distributed between $-a$ and $+a$ is alternately added to and subtracted from the bond energies on the hexagon. The move is rejected if any of the bond energies fall outside the interval $[-\frac{1}{3},\frac{2}{3}]$.}
\label{montecarlomove}
\end{figure}

While these single spin moves, in principle, are sufficient for correctly sampling the entire phase space, we also employed hexagon update moves to speed up the simulations at low temperatures. Given any configuration, we randomly choose a hexagon on the honeycomb lattice. To obtain the new configuration of spins we move along this hexagon, alternately adding and subtracting a quantity, $\Delta$, to the bond-energies, and then computing the spin components which give rise to these bond-energies. In Fig. \ref{montecarlomove}, suppose the topmost A-site is $s_1$, then $\epsilon_{1} = {S_{s_{1}}^{x}}^2 -\frac{1}{3}$ and $\epsilon_{2} = {S_{s_{1}}^{y}}^2 -\frac{1}{3}$. Now $\epsilon_{1}$ is changed to $\epsilon_{1} + \Delta$ and $\epsilon_{2}$ to $\epsilon_{2} - \Delta$. This then fixes the new ${S_{s_{1}}^{x}}$ and ${S_{s_{1}}^{y}}$ (up to a randomly chosen sign), leaving $S_{s_{1}}^{z}$ unchanged. Clearly, this leaves the sum of squares of the spin components unchanged. This change is also made to the four other bonds on the hexagon [Fig. \ref{montecarlomove}]. The value of $\Delta$ is chosen uniformly in the interval $[-a, a]$, where $a$ is a parameter. The proposed move is rejected if any of the bond-energies fall outside the interval $[-\frac{1}{3},\frac{2}{3}]$. Since the sum of bond-energies at each site (A and B) remains constant, these hexagon update moves leave the value of the effective Hamiltonian unchanged. These moves therefore play a crucial role in efficiently sampling the configurations close to the ground states at very low temperatures. We take the ratio of the phase space factors of the two states and accept or reject the proposed configuration according to the Metropolis rule. Clearly, this also satisfies the detailed balance condition. For the zero temperature simulations only the hexagonal updates were used.

Monte Carlo simulation data presented in this section has been computed for $L \times L$ triangular lattices of A-sublattice spins of various sizes, with $L$ ranging from $30$ to $256$. $6 \times 10^6$ Monte Carlo updates were made per site of which the first $6 \times 10^5$ were not used in computing the correlation functions. Correlation functions were calculated after every 6 updates per site.

\begin{figure}
\centering
\includegraphics[width=1.0\columnwidth,angle=0]{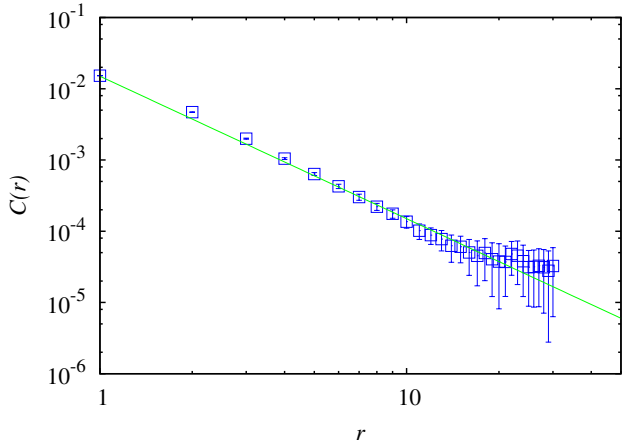}
\caption{(Color online) Plot of the zero temperature correlation function $C(\vec{r})=\langle {S_A^x}^2(0){S_A^x}^2(\vec{r})\rangle - \frac{1}{9}$ versus distance, $r$ along the $\hat{e}_{x}$ direction. These correlations follow a power law behaviour with exponent $ \simeq -2$. The line has a slope of $-2$.}
\label{corr11_zero}
\end{figure}

We calculated the correlation function $C(\vec{r})$ = $\langle{S_{a(l,m)}^z}^2 {S_{a(l',m')}^z}^2\rangle$ - $\frac{1}{9}$ for various lattice sizes and temperatures. Here $\vec r$ is the vector from site $a(l,m)$ to $a(l',m')$. We find that this correlation decays quite fast, and at finite temperatures, is very small except for a few points around the origin. At zero temperature, $C(\vec{r})$ is oscillatory along the $(1,0)$ direction (and is periodically negative). We observe a clear $\frac{1}{R^{2}}$ behaviour along the $\hat{e}_{z}$ direction, as plotted in Fig. \ref{corr11_zero}.

We looked for a signal of possible order in the ground state ensemble of the type proposed by 
Baskaran, et al.. One way of determining if there is any periodic order in the system is to study the structure factor which we define as the Fourier transform of $C(\vec{r})$
\begin{equation}
S(\vec{k})= \frac{1}{\sqrt{L M}}\sum_{(l',m')}{(\langle{S_{a(l,m)}^z}^2{S_{a(l',m')}^z}^2\rangle - \frac{1}{9})} \exp{(i \vec{k}.\vec{r})},
\end{equation}
where the summation is over all sites $a(l',m')$ for a fixed $a(l,m)$ with $\vec r$ as defined earlier.
\begin{figure}
\centering
\includegraphics[width=1.0\columnwidth,angle=0]{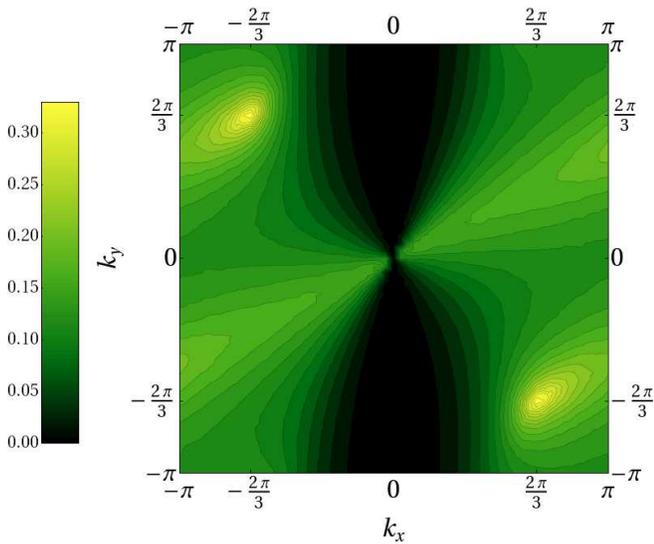}
\caption{(Color online) Plot of $\arrowvert S(\vec{k}) \arrowvert$, the ${s_A^z}^2$ structure factor, defined as $S(\vec{k})= \frac{1}{\sqrt{L M}} \sum_{\vec{r}}{(\langle{s_A^z}^2(0){s_A^z}^2(\vec{r})\rangle - \frac{1}{9})} \exp{(i \vec{k}.\vec{r})}$, where the $\vec{r}$ summation extends over all lattice sites. Two prominent peaks are visible at $(\frac{-2\pi}{3},\frac{2\pi}{3})$ and $(\frac{2\pi}{3},\frac{-2\pi}{3})$. However, these peaks do not diverge with system size in our simulations. }
\label{Qvsk}
\end{figure}
The structure factor $S(\vec{k})$ would have a delta function peak at $\vec{k}=(\frac{2 \pi}{3}, - \frac{2 \pi}{3})$ and $\vec{k}=(- \frac{2 \pi}{3}, \frac{2 \pi}{3})$, if there was an ordering of the type suggested in \cite{baskaran}. On calculating the structure factor for various $(k_{1},k_{2})$ at zero temperature, we find that $S(\vec{k})$, apart from some fluctuation all through, has two clearly visible peaks at wave vectors $(\frac{2\pi}{3}, -\frac{2\pi}{3})$ and $(-\frac{2\pi}{3},\frac{2\pi}{3})$, see Fig. \ref{Qvsk}. However the height of these peaks are only about three times the average value, and they do not become sharper with system size. Thus, we find no evidence of even incipient long-range order (hexatic-like, with power-law decay of the two-point correlation function) in the system at $T=0$.

\begin{figure}
\centering
\includegraphics[width=.7\columnwidth,angle=270]{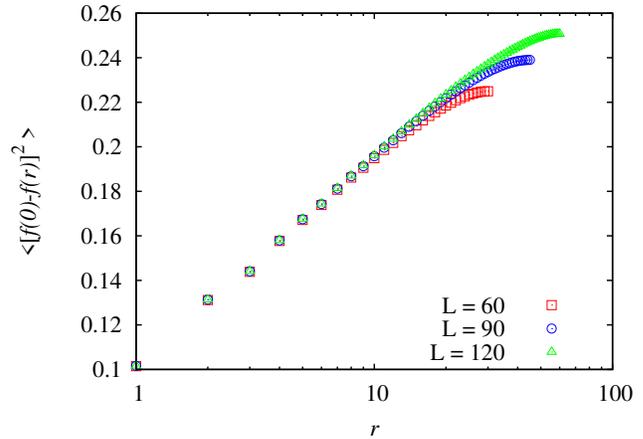}
\caption{(Color online) Graph of the zero temperature correlation function $\langle(f(0)-f(r))^2\rangle$ versus distance, $r$, for different lattice sizes $L$, showing a $\log r$ dependence in accordance with the mapping to a height model at a finite temperature.}
\label{ffzero}
\end{figure}

\begin{figure}
\centering
\includegraphics[width=.7\columnwidth,angle=270]{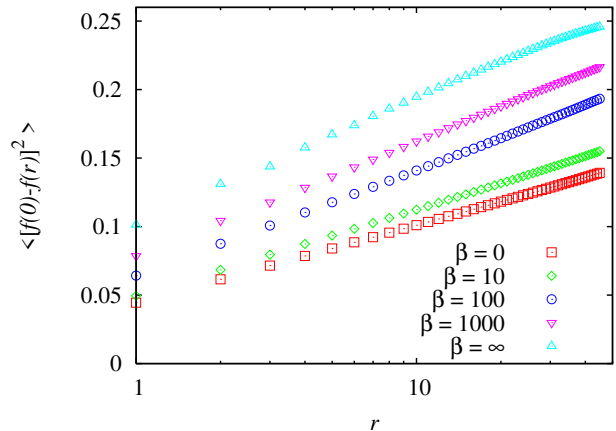}
\caption{(Color online) Graph showing the finite temperature correlation function $\langle(f(0)-f(r))^2\rangle$ versus distance, $r$ for various values of $\beta = T^{-1}$. The correlations are logarithmic at all temperatures, with the coefficient of $\textmd{log}(r)$ varying between $(2.45 \pm 0.05) \times 10^{-2} \simeq \frac{2 \sqrt{3}}{45 \pi}$ at $\beta = 0$ and $(4.12 \pm 0.05) \times 10^{-2}$ at $\beta = \infty$. }
\label{ffbeta}
\end{figure}

We also computed correlations of the $f$ and $\phi$ fields at various temperatures. At the end of each Monte Carlo step the $\phi$ field was generated from the spin configuration by solving the discrete Poisson equation on the triangular lattice (Eq. (\ref{discrete_poisson})). This was done by inverting the Poisson equation in Fourier space as shown in Eq.(\ref{inverse_fourier}). The Fourier transforms were calculated using the fast Fourier transform code provided in \cite{numerical_recipes}. The spin configuration and the $\phi$ field was then used to generate the $f$-field using Eq.(\ref{eq:def-epsilon}). 
In Fig. \ref{ffzero}, we have plotted the zero temperature correlation function $\langle (f_{l,m} - f_{l',m'})^2 \rangle \equiv \langle(f(0)-f(r))^2\rangle$ versus $ \log r$ where $r$ is the distance between the two sites. We see that this correlation function increases logarithmically with distance. Note that a logarithmic dependence of this function implies a $1/r^2$ dependence of the bond-energy bond-energy correlation function (Eq. (\ref{powerlaw})). Fig. {\ref{ffbeta}} shows the correlation function $\langle(f(0)-f(r))^2\rangle$ at vaious values of $\beta$. These correlations vary as log$(r)$ at all temperatures with the coefficient varying between $(2.45 \pm 0.05) \times 10^{-2} \simeq \frac{2 \sqrt{3}}{45 \pi}$ at $\beta = 0$ and $(4.12 \pm 0.05) \times 10^{-2}$ at $\beta = \infty$. 

\section{The large-S quantum Kitaev model}

In this section we would like to discuss the ground-state energy of quantum spin-$S$ Kitaev model using a variational approach, which does not suffer from divergences. We obtain upper and lower bounds on the ground state energy. Our quantum mechanical Hamiltonian is normalised by the size of the spins. So 

\begin{align}
\nonumber
& H = - \frac{J}{S(S+1)} \sum_{a \in A } [ S_{a}^{x} S_{a+e_x}^{x} + S_{a}^{y} S_{a+e_y}^{y} 
&  + S_{a}^{z} S_{a+e_z}^{z} ] .
\label{def_Hamiltonian}
\end{align}

Using the operator inequality $ AB \geq -( A^2 + B^2)/2$, where $A$ and $B$ are any commuting Hermitian operators, it is easily seen that $H$ satisfies the lower bound 

\begin{equation}
E_{ground} \geq -JN.
\label{trivial_bound}
\end{equation}

A better bound may be proved as follows. We write $H$ as

\begin{equation}
H = \frac{1}{S(S+1)}\sum_{(l,m)} H_{b(l,m)},
\end{equation}
where $H_{b(l,m)}$ is a 4-site spin Hamiltonian containing only the couplings of the site $b(l,m)$ on the B-sublattice and its neighbours

\begin{align}
\nonumber
H_{b(l,m)} = & -J(\hat S_{a(l,m+1)}^{x} \hat S_{b(l,m)}^{x}+ \hat S_{a(l-1,m+1)}^{y} \hat S_{b(l,m)}^{y} \\
 & + \hat S_{a(l,m)}^{z} \hat S_{b(l,m)}^{z}).
\end{align}

The operators $H_{b(l,m)}$ can be diagonalized in a Hilbert space of 4-spins, i.e. a $(2S+1)^4$ 
dimensional Hilbert space. We note that 
$ \hat{S}_{a(l,m+1)}^{x}, \hat S_{a(l-1,m+1)}^{y}, \hat S_{a(l,m)}^{z}$ are operators that belong to different sites, they commute amongst each other and with $H_{b(l,m)}$. 
Hence we can move to a basis in which these are diagonal. We now work in the subspace in which the 
basis vectors are eigenvectors of $\hat{S}_{a(l,m+1)}^{x}, \hat S_{a(l-1,m+1)}^{y}$ and 
$\hat S_{a(l,m)}^{z}$ with eigenvalues $ {s}_{a(l,m+1)}^{x}, s_{a(l-1,m+1)}^{y}, s_{a(l,m)}^{z}$ respectively. Thus
the eigenvalues $\lambda$ of $H_{b(l,m)}$ satisfy the relation 

\begin{equation}
\lambda^2 \leq
 (JS)^{2} ({s_{a(l,m+1)}^{x}}^{2} + {s_{a(l-1,m+1)}^{y}}^{2} + {s_{a(l,m)}^{z}}^{2}).
\end{equation} 

This is true for all eigenvalues $ {s}_{a(l,m+1)}^{x}$, $s_{a(l-1,m+1)}^{y}$, $s_{a(l,m)}^{z}$ and hence is valid as an operator inequality 

\begin{align} 
&H_{b(l,m)}^2 \leq\\ 
&(JS)^2 \left[ (\hat{S}^{x}_{a(l,m+1)})^2 + (\hat S_{a(l-1,m+1)}^{y})^2 + (\hat S_{a(l,m)}^{z})^2 \right].
\end{align}
Therefore, for any wavefunction $|\psi \rangle$ of all the $2N$ spins on the lattice we have

\begin{align} 
 \nonumber
& \langle\psi|H_{b(l,m)}^2|\psi\rangle \leq \\
& (JS)^2 \left[ \langle(\hat{S}_{a(l,m+1)}^{x})^2\rangle + \langle(\hat S_{a(l-1,m+1)}^{y})^2\rangle + \langle(\hat S_{a(l,m)}^{z})^2\rangle \right]
\end{align}
where $\langle(\hat{S}_{a(l,m+1)}^{x})^2\rangle = \langle \psi|(\hat{S}_{a(l,m+1)}^{x})^2|\psi\rangle$ and so on. Using the fact that $\langle\psi|H_{b(l,m)}|\psi\rangle^2 \leq \langle\psi|H_{b(l,m)}^2|\psi\rangle$ and taking the square root we get
{\small
\begin{align}
\nonumber
& \langle\psi|H_{b(l,m)}|\psi\rangle \\
& \geq -J S \sqrt{\langle(\hat{S}_{a(l,m+1)}^{x})^2 \rangle + \langle(\hat S_{a(l-1,m+1)}^{y})^2 \rangle + \langle(\hat S_{a(l,m)}^{z})^2\rangle}.
\end{align}
}
This immediately gives
\begin{align}
\nonumber
& \langle\psi|H|\psi\rangle \geq \sum_{(l,m)}\\
& -J S \sqrt{\langle(\hat{S}_{a(l,m+1)}^{x})^2 \rangle + \langle(\hat S_{a(l-1,m+1)}^{y})^2 \rangle + \langle(\hat S_{a(l,m)}^{z})^2\rangle},
\label{quantum_sqrt_sum}
\end{align}
where the sum is over all the sites of the B-sublattice. Note that the terms in each of the squareroots are all real numbers.

Using equation (\ref{convex_ineq}) in the equation (\ref{quantum_sqrt_sum}) and observing that for any site on the A-sublattice $(\hat{S}_{a(l,m)}^{x})^2 + (\hat S_{a(l,m)}^{y})^2 + (\hat S_{a(l,m)}^{z})^2 = S(S+1)$ we get

\begin{equation}
\langle\psi|H|\psi\rangle \geq -J N \sqrt{\frac{S}{S+1}}.
\end{equation}
For large $S$, 
\begin{equation}
\frac{E_G}{JN} \geq -\frac{1}{2} + \frac{1}{4 S} + O(\frac{1}{S^2})
\end{equation}
is a lower bound, which shows an increase in the ground state energy due to quantum fluctuations. 

We now describe a variational upper bound for the ground state energy of the quantum model.

We take a trial wave function which is a direct product of two-site pair wavefunctions:
\begin{equation}
|\psi\rangle = \Pi_{(l,m)} |\psi_{a(l,m) b(l,m)}\rangle.
\end{equation}
Each of these two wavefunctions $|\psi_{a(l,m) b(l,m)}\rangle$ is the ground state wavefunction of the two site Hamiltonian $\mathcal{H}_2$ given by 

\begin{equation}
\mathcal{H}_2 (\lambda) = - [\lambda (S^x_{a(l,m)} + S^x_{b(l,m)}) + S^z_{a(l,m)} S^z_{b(l,m)}]
\label{H2}.
\end{equation}
Using $\lambda$ as the variational parameter we calculate the resulting ground state energy of the full Hamiltonian. 

Let $\gamma^{x}(\lambda)$ = $\langle S^{x}_{a(l,m)} \rangle_{\mathcal{H}_2}$, and $e^{zz}(\lambda) = \langle S^z_{a(l,m)} S^{z}_{b(l,m)} \rangle_{\mathcal{H}_2}$, where $\langle ~ \rangle _{\mathcal{H}_2}$ denotes average under the ground state of $\mathcal{H}_2$ [Eq. (\ref{H2})]. Let $E(\lambda)$ be 
the minimum eigenvalue of $\mathcal{H}_2$. Then, clearly for the wavefunction $|\psi\rangle$, we have 
\begin{equation}
\langle \psi| H | \psi \rangle = - N J [ e^{zz}(\lambda) + \gamma^x(\lambda)^2].
\end{equation}

\begin{figure}
\centering
\includegraphics[width=1.0\columnwidth,angle=0]{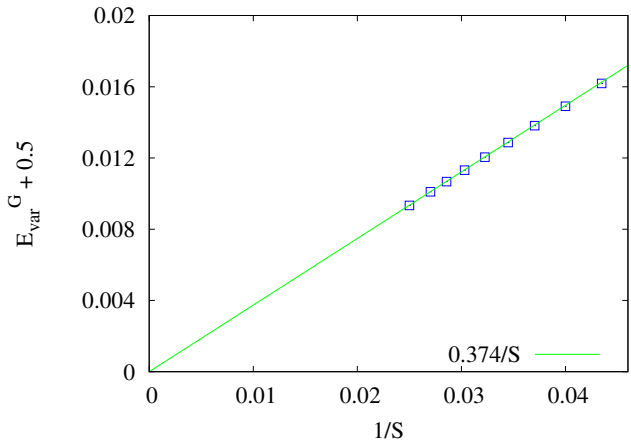}
\caption{(Color online) Plot of $E^G_{var} +0.5$ as a function of $\frac{1}{S}$ displaying a slope of $0.374 \pm 0.005$.}
\label{quant_var}
\end{figure}

It is straightforward to determine the minimum eigenvalue $E_2(\lambda)$ by numerical diagonalization of the corresponding $(2S +1)^2 \times (2S+1)^2$ dimensional matrix for different values of $\lambda$. We can then also determine $\gamma(\lambda)$ and $e^{zz}(\lambda)$ numerically from the corresponding eigenvector. The resulting upper bound on the ground state energy per site $E^G_{var}$
was determined for different values of $S$ by minimizing over $\lambda$. Fig. (\ref{quant_var}) shows a plot of $(E^G_{var} +0.5)$ versus $\frac{1}{S}$. We see that the energy excess over the classical ground state energy varies as $1/S$ for large $S$, 
\begin{equation}
\frac{E_G}{JN} \leq E^G_{var} = - \frac{1}{2} + 0.374 \frac{1}{S} + O(\frac{1}{S^2}).
\end{equation}
The value of the coefficient of the $1/S$ correction term obtained by us ($0.374$) should be compared with the estimate $0.289$ by Baskaran, et al. \cite{baskaran} using the quadratic approximation. The latter underestimates the true answer, as the modes that have frequency zero in the quadratic approximation would actually have a finite contribution. In our calculation, these
corrections are not ignored. Also, the variational estimate can be improved with better choice of trial wavefunction $|\psi\rangle$, say, in terms of six-site clusters.

The ground state manifold of the classical problem has several soft modes. This suggests that one can find local ground states such that forming a product state with these as a basis yields a good approximation to the ground state of the full system. Then forming a linear superposition of such states locally, and forming products, lifts the degeneracy partially and gives a better variational ground state. We perform such a calculation as follows.

The sites of the hexagonal lattice can be divided into disjoint hexagons. Thus the full Kitaev model Hamiltonian consists of interaction terms between sites within the same hexagon, and between different hexagons. We write $H = \sum_{hex} H_{hex} + \sum_{int} H_{int}$, where the second sum is over the interconnecting bonds. Now some of the interconnecting bonds are of X-type, some Y-type and some Z-type. We can redefine the directions $x-$ , $y-$ and $z-$ at each site so that all the interconnecting bonds are of Z-type. Then the couplings between sites within a hexagon are alternating X- and Y-type, and the Hamiltonian $H_{hex}$ for a single hexagon, with sites labelled $1,2, \ldots 6$ is of the form 

\begin{align}
\nonumber
H_{hex} = & -\frac{J}{S(S+1)} [ S^x_1 S^x_2 + S^y_2 S^y_3 + S^x_3 S^x_4 + S^y_4 S^y_5 \\
 & + S^x_5 S^x_6 + S^y_6 S^y_1 ].
\label{clus_Hamiltonian}
\end{align}

For the classical Hamiltonian $H_{hex}$, a state in which all spins are aligned in the same direction $\phi$ in the $xy-$ plane are classical ground states. The quantum state corresponding to this state denoted by $|\phi \phi \phi \phi \phi \phi\rangle$ may be written as a product of single-site coherent states at the six sites
\begin{equation}
|\phi \phi \phi \phi \phi \phi\rangle = |\phi\rangle_1 \otimes |\phi\rangle_2 \otimes |\phi\rangle_3 \otimes |\phi\rangle_4 \otimes |\phi\rangle_5 \otimes |\phi\rangle_6,
\end{equation}
where $|\phi\rangle_j$ represents the wavefunction of the spin at site $j$, polarized in the direction $\phi$ in the x-y plane, and can be written as 
\begin{equation}
|\phi\rangle_j = \left( \frac{1}{2} \right)^S e^{(e^{\iota \phi} \hat{S}^-_j)} |S\rangle_j
\label{coh_st},
\end{equation}
where $S^- \equiv S^x - \iota S^y$ is the angular momentum lowering operator at the site $j$, and $|S\rangle_j$ is the state with $S^z_j = S$.

Since states for different $\phi$ are classically degenerate, a better quantum-mechanical trial state is a linear superposition of these states 

\begin{equation}
|\psi_1\rangle = \frac{1}{2 \pi} \int_0^{2 \pi} d \phi e^{-\iota 6S \phi} 
|\phi \phi \phi \phi \phi \phi\rangle.
\end{equation}

The full wavefunction is a direct product of wavefunctions for different hexagons. The matrix elements of the spin operators are easy to calculate \cite{radcliffe}. In this state, the expectation value of $H_{int}$ vanishes. Therefore we have
\begin{equation}
E_{G} \leq \frac{1}{6} \frac{\langle\psi_1|H_{hex}|\psi_1\rangle}{\langle\psi_1|\psi_1\rangle}
\end{equation}
The expectation value of $ H_{hex} $ is easily seen to be

\begin{equation} \frac{\langle\psi_1|H_{hex}|\psi_1\rangle}{\langle\psi_1|\psi_1\rangle} = -3 J \left[ 1 -\frac{11}{12S} +\ldots \right] ,
\end{equation}
which is a bit better than the variational bound for energy $-3 J \frac{S}{S+1}$, using the state $|\phi\phi\phi\phi\phi\phi\rangle$.

\section{Discussion}

We have shown that the Kitaev model with classical spins shows no order-by-disorder, while there are plausible arguments that the quantum model does. This is because the mechanism of order -by- disorder in classical and quantum models is somewhat different.

Consider a classical model, whose ground states form an M-dimensional manifold. Let $G$ be one of the ground states. We expand the energy in the coordinates orthogonal to the manifold, and look for small perturbations about the ground state. Keeping terms in the deviations from the ground state to quadratic order, and going into the normal mode coordinates, we get a quadratic approximation to the Hamiltonian in the transverse coordinates as 

\begin{equation}
\delta H = \sum_{j} [ \frac{1}{2m_j(G) } p_j^2 + \frac{1}{2}m_{j}(G)\omega_j(G) ^2 q_j^2 ] ,
\end{equation}
where the sum over $j$ extends over the $n$ transverse degrees of freedom. 
Then, the corresponding quantum mechanical partition function in the quadratic approximation is 

\begin{equation}
Z_{quad} \approx \prod_{j} \frac{e^{-\beta \omega_j/2}}{ 1 - e^{-\beta \omega_j}}.
\end{equation}

For low temperatures, the $G$ having the minimum value of the effective quantum mechanical free energy is obtained by minimizing the ``zero-point energy'' $f_{q}(G) = \sum_j\frac{1}{2} \hbar \omega_j$.
However, the classical partition function corresponds to the case $ \beta \omega_j \ll 1$, and at low temperature $T$ is easily seen to be proportional to $ T^n / \prod{ \omega_j (G)}$. Thus the relative weights of different points $G$ on the manifold are determined by an effective free energy $ f_{cl}(G)$ 
proportional to $ \sum_{j} \log \omega_j$. Clearly $f_q$ and $f_{cl}$ are quite different, and states which are favored by one need not be favored by the second. In particular, $f_{cl}$ depends more sensitively on the low frequency modes.

If some of the $\omega_j$'s are zero, in this approximation, the classical partition function diverges, but the quantum weight has no singularity. This problem of zero modes also occurs in the calculation of \cite{baskaran}. In fact the zero frequency eigenvalue has a large degeneracy.

More generally, finite $\hbar$ corrections in a quantum mechanical system correspond to a finite temperature classical model but in one higher dimension, and can be qualitatively different. A simple example of this is a system of masses coupled by nearest neighbour springs in one dimension. In the classical case, the variance of displacements of masses at distance $R$ varies as $T R$ for large $R$, and small $T$, but this quantity grows only as $\log R$, both in the quantum case at zero temperature, and the classical case in {\it two } dimensions. 

In the path-integral formulation, the large-$S$ quantum Kitaev model becomes a set of classical
spins on a $2+1$ dimensional lattice, with Kitaev couplings in two spatial directions, and ferromagnetic couplings in the time/inverse-temperature direction. It is quite plausible that in this $3$-dimensional model, there is long-range order for low ``effective temperature", but in the $2$-dimensional classical Kitaev model, the destablizing effect of fluctuations is too strong.  

There are several interesting classical 2D systems, where thermal Order-by-Disorder is expected. The prototypical example is the system of Heisenberg spins on a kagome lattice, with nearest neighbour antiferromagnetic couplings. The expectation of order-by-disorder in this classical system comes from theoretical and Monte-Carlo studies, that suggest that at low temperatures, the spins lie on a single plane as $T \rightarrow 0$ \cite{chalkerholdsworth},\cite{zhitomirsky}. This model can also be related to the height model at its critical point, within the quadratic approximation, suggesting that a single long-range ordered state (the $\sqrt{3}\times \sqrt{3}$ state) is selected over the coplanar ones \cite{huserutenberg},\cite{henley4}. It would be interesting to identify the main reason for the difference in the behaviors in these models and the case studied here.

We have mapped the finite and zero temperature problem of classical spins onto a height-model interacting via an effective Hamiltonian $H_{SOS}$. This effective Hamiltonian depends on the temperature of the spin model. We have shown that the correlations of the height variables vary as $\textmd{log}(r)$, where $r$ is the distance between the sites, for all temperatures of the spin model. The discrete height model with a pinning potential in 2D undergoes a roughening transition from a phase in which it is ordered to one with logarithmic correlations between the height variables \cite{Jose}. In the rough phase ($T > T_{R}$) of the height model, the coefficient of $\textmd{log}(r)$ gives us a measure of the temperature \cite{Chiu_Weeks},\cite{Ohta}. So we see that the range of temperature $[0,\infty]$ of the spin model maps onto a range of temperature of the height model which is in the rough phase. Note that as the temperature of the Kitaev Hamiltonian is decreased, $\langle(f(0)-f(r))^2\rangle$ in the SOS model increases. The increased fluctuations in $f$ are accompanied by a decrease in fluctuations of the $\phi$ field, and the fluctuations of the spins $\{\vec{S}\}$ decrease with temperature, as expected.

\begin{acknowledgements}
We thank G. Baskaran, R. Shankar, Diptiman Sen, Kedar Damle and M. Mezard for very useful discussions. DD acknowledges support of the Government of India through J. C. Bose fellowship. SC thanks CSIR, Government of India for financial support.

\appendix
\section{$f-f$ Correlation functions at Infinite Temperature}
In this Appendix we calculate the asymptotic behaviour of the $f-f$ correlation functions at infinite temperature.
There are two independent degrees of freedom at each A-site. We can choose these to be $\epsilon (l,m; z)$ and $\epsilon (l,m; y)$. Then $\epsilon (l,m; x)$ is clearly $-\epsilon (l,m; z)-\epsilon (l,m; y)$. At infinite temperature

\begin{align}
\nonumber
 \langle\epsilon (l,m; z) \epsilon (l',m'; z)\rangle = \frac{4}{45}\delta_{l,l'}\delta_{m,m'},\\
 \langle\epsilon (l,m; y) \epsilon (l',m'; z)\rangle = \frac{-2}{45}\delta_{l,l'}\delta_{m,m'}.
\label{infinitetempcorrelations}
\end{align}
 
For a field $\phi(l,m)$ on the lattice, we define the Fourier and inverse Fourier transforms as follows 
\begin{align}
\nonumber
\phi (\vec{k}) = \frac{1}{\sqrt{LM}}\sum_{\vec{r}} \textmd{exp}[i \vec{k}.\vec{r}]\phi (\vec{r}),\\
\phi (\vec{r}) = \frac{1}{\sqrt{LM}}\sum_{\vec{k}} \textmd{exp}[-i \vec{k}.\vec{r}]\phi (\vec{k}).
\label{transform_definition}
\end{align}
The vector $\vec{r}$ denotes the point $(l,m)$ in real space and $\vec{k} \equiv (k_{1},k_{2})$ denotes the point $(\frac{u}{2 \pi L},\frac{v}{2 \pi M})$ in Fourier space, with $\vec{k}.\vec{r} = \frac{u l}{2 \pi L} + \frac{v m}{2 \pi M}$. The charge $Q_{b(l,m)}$ at each B-sublattice site is given by 
\begin{align}
\nonumber
Q_{b(l,m)} =-\epsilon (l,m; z) - \epsilon (l-1,m+1; y) \\
+ \epsilon (l,m+1; z) + \epsilon (l,m+1; y).
\label{charge_lm}
\end{align}

The discrete Poisson equation that determines the potential fields $\phi$ is 
\begin{align}
\nonumber
 \phi_{a}(l,m)+\phi_{a}(l-1,m+1)+\phi_{a}(l,m+1)-3\phi_{b}(l,m),\\
\nonumber
=-Q_{b(l,m)}\\
\nonumber
 \phi_{b}(l,m)+\phi_{a}(l+1,m-1)+\phi_{a}(l,m-1)-3\phi_{a}(l,m),\\
=0
\label{discrete_poisson}
\end{align}
where $\phi_{a}(l,m)\equiv \phi(a(l,m)) $ and $\phi_{b}(l,m)\equiv \phi(a(l,m) + e_{z})$.
Eq.(\ref{discrete_poisson}) can be inverted in Fourier space as
\begin{equation}
\left[\begin{array}{c}
\phi_{a}(\vec{k}) \\ 
\phi_{b}(\vec{k})\\
\end{array}\right] = \frac{-1}{9-h(\vec{k})h^{*}(\vec{k})}\left[\begin{array}{cc}
3 ~~~~~h(\vec{k})\\ 
h^{*}(\vec{k}) ~~~~3\\
\end{array}\right]
\left[\begin{array}{c}
0\\ 
-Q_{b}(\vec{k})\\
\end{array}\right],
\label{inverse_fourier}
\end{equation}
where $h(\vec{k}) = 1 + \textmd{exp}[i(k_{2} -k_{1})] + \textmd{exp}[i k_{2}]$.
Now, the difference in the $f$ variables along the z axis is given by
\begin{align}
\nonumber
f(l+R,m+1)-f(l,m+1)\equiv f_{R}-f_{0} = \\
 \sum_{r=1}^{R} \epsilon(l+r,m;z) + \sum_{r=1}^{R} [\phi_{b}(l+r,m)-\phi_{a}(l+r,m)].
\end{align}
We can write this in terms of the Fourier components as follows (taking $(l,m)=(0,0)$):

{\small
\begin{equation}
 \begin{split}
 &[f_{R}-f_{0}] = \frac{1}{L M}\sum_{r=1}^{R} \sum_{\vec{k}} 
 [\alpha_{\vec{k}} \epsilon(\vec{k};z) + \beta_{\vec{k}} \epsilon(\vec{k};y)] 
 \textmd{exp}[-i k_{1} r],\\
 &\textmd{where} ~~\alpha_{\vec{k}}=
 1+\frac{h(\vec{k})-3}{9-h(\vec{k}){h^*(\vec{k})}}(1-\textmd{exp}[-ik_{2}])\\
&\textmd{and} ~~\beta_{\vec{k}}=  \frac{h(\vec{k})-3}{9-h(\vec{k})h^{*}(\vec{k})}(\textmd{exp}[ik_{1}]-1)(\textmd{exp}[-ik_{2}]).
 \end{split}
\end{equation}
}

Summing over $r$ first and using Eq. (\ref{infinitetempcorrelations})
\begin{align}
\nonumber
{\langle(f_{R}-f_{0})^{2}\rangle}_{\beta = 0} = \frac{1}{L M} \sum_{\vec{k}} {\left| \frac{1-\textmd{exp}[-ik_{1} (R+1)]}{1-\textmd{exp}[-ik_{1} ]}\right|}^{2}\times\\
[\langle {(\alpha_{\vec{k}} \epsilon(\vec{k};z) + \beta_{\vec{k}} \epsilon(\vec{k};y))}^{2} \rangle].
\end{align}
The term within the square brackets can be shown to be equal to $ \frac{6}{45}(\frac{1-\cos k_{1}}{3-\cos k_{1} -
\cos k_{2}-\cos(k_{1}-k_{2})})$. Thus the correlation function simplifies to
{\small
\begin{equation}
\begin{split}
&{\langle(f_{R}-f_{0})^{2}\rangle}_{\beta = 0} = \\
& \frac{6}{45}\frac{1}{L M}\sum_{\vec{k}}{\frac{1- \cos(k_{1}(R+1))} {3- \cos k_{1} -
\cos k_{2}- \cos (k_{1}-k_{2})}}.
\end{split}
\end{equation}
}
 We note that the factors of $1-\cos{k_1}$ cancel out in the expression. In the limit $L,M \rightarrow \infty$ this summation becomes an integral giving
\begin{equation}
{\langle(f_{R}-f_{0})^{2}\rangle}_{\beta=0} = \frac{6}{45}G(0,R),
\end{equation}
where $G(0,R)$ is the lattice Green's function on the triangular lattice between the points $(0,0)$ and $(0,R)$ which is equal to $\frac{1}{ \pi \sqrt{3}}\textmd{log}[R]$ at large $R$ \cite{Horiguchi}. Thus we have
\begin{equation}
\langle(f_{R}-f_{0})^{2}\rangle_{\beta=0} = \frac{2 \sqrt{3}}{45 \pi} \textmd{log}[R] + {\cal O}(1) ~~~~\textmd{for large R}.
\end{equation}

\end{acknowledgements}

\end{document}